# Designing a broadband and linear polarization metasurface carpet cloak in the visible

L.Y. Hsu[1,4], A. Ndao[1,4], and B. Kanté[1,2,3,4]

[1] Department of Electrical Engineering and Computer Sciences, University of California, Berkeley, California 94720, United States

[2] Department of Mechanical Engineering, University of California, Berkeley, California 94720, United States

[3] Materials Sciences Division, Lawrence Berkeley National Laboratory, 1 Cyclotron Road, Berkeley, California 94720, United States

[4] Department of Electrical and Computer Engineering, University of California San Diego, La Jolla, California 92093-0407, United States

bkante@berkeley.edu

**Abstract:**

*In the past few years, carpet cloaking attracted interests because of its feasibility at optical frequencies and potential in stealth technologies. Metasurfaces have been proposed as a method to engineer ultra-thin carpet cloaking surfaces due to their abilities to manipulate wavefronts, polarization, and phase at subwavelength scale. However, achieving broadband carpet cloaking with a significant bandwidth is one of the key remaining challenges for metasurface designs. To date, broadband carpet cloaking based on metasurfaces has not been achieved and cloaking is limited to discrete wavelengths. Here, we propose and numerically demonstrate a novel metasurface design for broadband carpet cloaking with linear polarization at visible wavelengths from 650 nm to 800 nm. Our proposed method is a promising approach for broadband structured interfaces.*

**Introduction**

In the past few years, metasurfaces have been investigated as potential alternatives for bulky integrated optical free space components due to the simplicity of their manufacturing. Metasurfaces are subwavelength nanostructured devices that enable the control of optical wavefronts, polarization, and phase. A large variety of flat optical components, including planar lenses [1-13], solar concentrators [14], polarizers [15], thin absorbers [16-17], sensors [18-19], or carpet cloaks [20-27] have been demonstrated. However, the working bandwidth of currents metasurfaces, despite significant effort, are limited by not only the intrinsic optical properties of materials in nature but also their design principle. Although, multiple wavelengths and relatively broadband carpet cloaking have been recently reported in [28-30], the proposed devices are so far limited to circular polarization, discrete wavelength, continues bandwidth from 1200 nm to 1680 nm in infrared [31].

Controlling material dispersion has been recognized as a linchpin to broadening optical bandwidth of components. The ability to control the dispersion is, thus, a highly desired property of the medium supporting optical modes. Therefore, minimizing the dispersion of material is the main concern in optical components specifically for broadband carpet cloaking. Conventional metasurfaces proposed for large band operation are based on optical waveguides made of a high-index core surrounded by low-index claddings which limit currents metasurfaces applications to relatively small bandwidth governed by the dispersion of the waveguides. To overcome these limitations, we propose an alternative solution that consists in using slot waveguides as they can confine light in the lossless material (air) compared to systems guiding light in high index materials, usually dispersive at higher frequencies.

Here, we develop a new approach based on slot-waveguide to mitigate material loss by engineering a low loss mode confined in air gaps. By using appropriate geometrical parameters, we designed and simulated a broadband metasurface carpet cloak with linear polarization in the visible range from 650 nm to 800 nm. To our knowledge, this is the first time such broadband metasurface carpet cloak with linear polarization is proposed in the visible. This new design enhances the capability of metasurfaces. The presented results thus pave the way to broadband metasurfaces for various applications and devices.

**Design Principle**

The carpet surface under which cloaking is achieved can be described by a surface z(x,y). To illustrate the design strategy, we consider a bump surface (invariant in y) that is described by a Gaussian function. An arbitrary shaped object can always be hidden below a surface of the this form.

$$z(x) = Ae^{-\frac{x^2}{\sigma^2}} \tag{1}$$

To achieve the cloaking function, the phase distribution required $\Phi(x)$ to compensate the phase advance/delay due to the bump is given by [21]

$$\begin{aligned}\Phi(x,f) &= -2k_0 z(x)\cos\theta_G + const(f) = -2\frac{2\pi}{c}z(x)\cos\theta_G f + const(f) \\ &= m(x)f + cont(f), m(x) = -\frac{4\pi}{c}z(x)\cos\theta_G\end{aligned} \tag{2}$$

where *const(f)* is defined by the flat surface, *f* is frequency of the incident wave, x is the position on bump and m is the slope of phase equation with respect to frequency *f*. By providing the appropriate phase distribution in equation 2, we can make the bump look like a flat ground plane using all dielectric metasurfaces. Equation 2 implies that a cloaking device can be broadband if the phase distribution on the metasurface is linear with respect to frequency (*f*). Secondly, the slope is different for different position (x) but const(*f*) is the same for different positions while it can be different for different frequency (*f*). However, because an absolute phase is meaningless in this system, only phase differences matter, an arbitrary phase can be chosen as phase reference. To simplify the equation, we choose the phase reference as cont(f), then the phase-shift ($\Phi_s$) distribution required for a broadband metasurface cloaking becomes:

$$\Phi_s(x,f) = \Phi(x,f) - cont(f) = -2k_0 z(x)\cos\theta_G = -2\frac{2\pi}{c}z(x)\cos\theta_G f = m(x)f \tag{3}$$

To demonstrate broadband carpet cloaking, we design a metasurface carpet cloak for a Gaussian bump where A = 360 nm, $\sigma$ = 2.4 μm, from 375 THz to 460 THz for an incident angle ($\theta_G$) 45°. Figure 1(a) presents 2D plot for phase-shift ($\Phi_s$) with x as x-axis and *f* as y-axis. Figure 1(b) presents phase-shift profiles for different frequencies. They are bump-shape curves and the phase-shift gradient with position increases with frequency. Figure 1(c) present phase-shift frequency profiles at different positions (x). They are linear with respect to the frequency and the phase-shift gradient increases with frequency as well as with position.

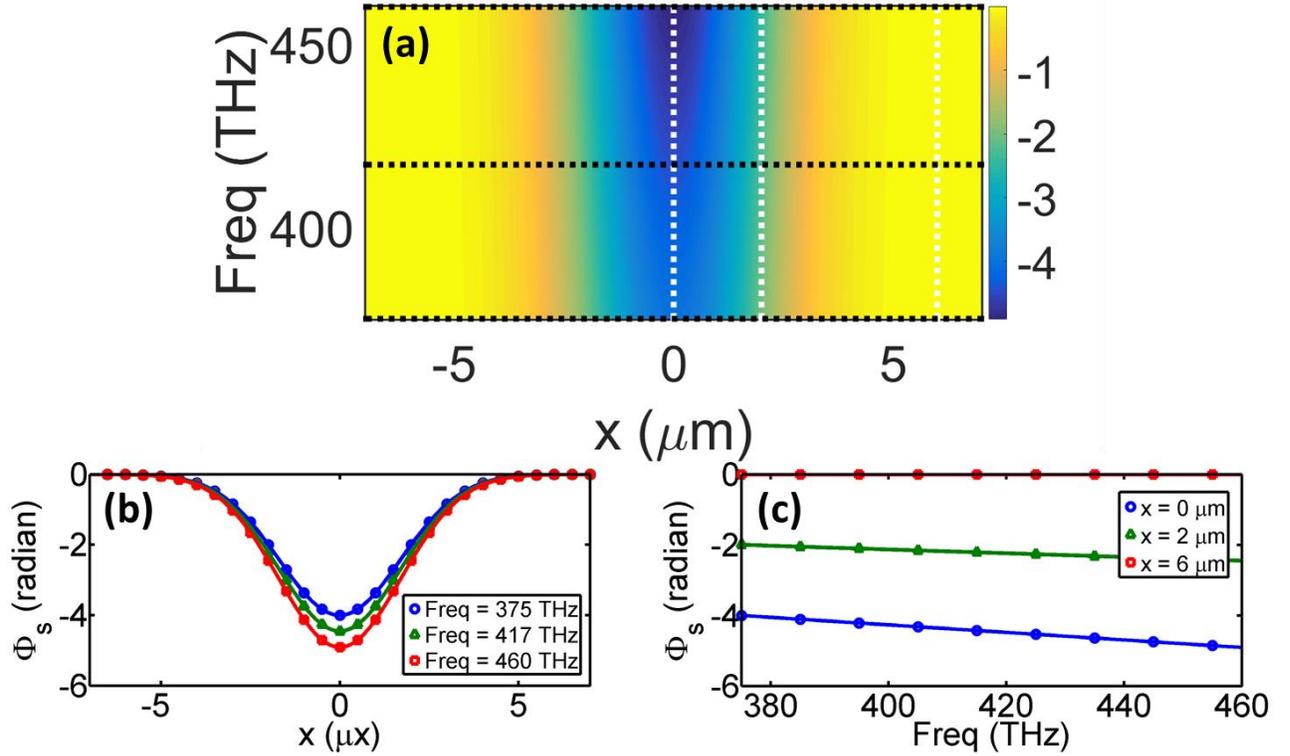

Fig. 1. (a) Required phase-shift profile (in radians) for a broadband metasurface carpet cloaking. (b) Phase-shift with position. (c) Phase-shift with frequency.

To design a broadband carpet cloak, we propose to use a slot waveguide as unit-cell in order to provide a linear phase with respect to frequency. The slot waveguide consists of two high index ridges ($Si_3N_4$), separated by a narrow low index gap (air). Due to such refractive index discontinuity, this structure allows one of the propagating modes to confine its energy within the slot region. The principle of operation of this structure is based on the discontinuity of the electric field at the high-index-contrast interface [32]. The advantage of using slot waveguides as a unit cells is to mitigate the material dispersion by confining the mode in air gap and enables subwavelength confinement of the light.

Figure 2(a) presents the schematic of the carpet cloaking design. The incident wave is TM polarized wth the magnetic field H field along the y direction and with an incident angle $\theta_G$ presented on the left. Figure 2(b) presents the geometry of the unit cell of the metasurface. The period (p) of the metasurface array is 280 nm, the width (w) of the air gap varies from 80nm to 200nm. The metal layer is silver with a thickness of 100 nm ($h_{Silver}$), the spacer is $SiO_2$ with thickness 350 nm ($h_{SiO_2}$), and the slot waveguide consists of two ridges of $Si_3N_4$ with a thickness 350 nm ($h_{Si_3N_4}$) separated by an air gap. Figure 2(c) shows the normalized electrical field at 409 THz with w = 80 nm and $\theta_L = 45°$. One can observe that the highest field confinement is in the air gap.

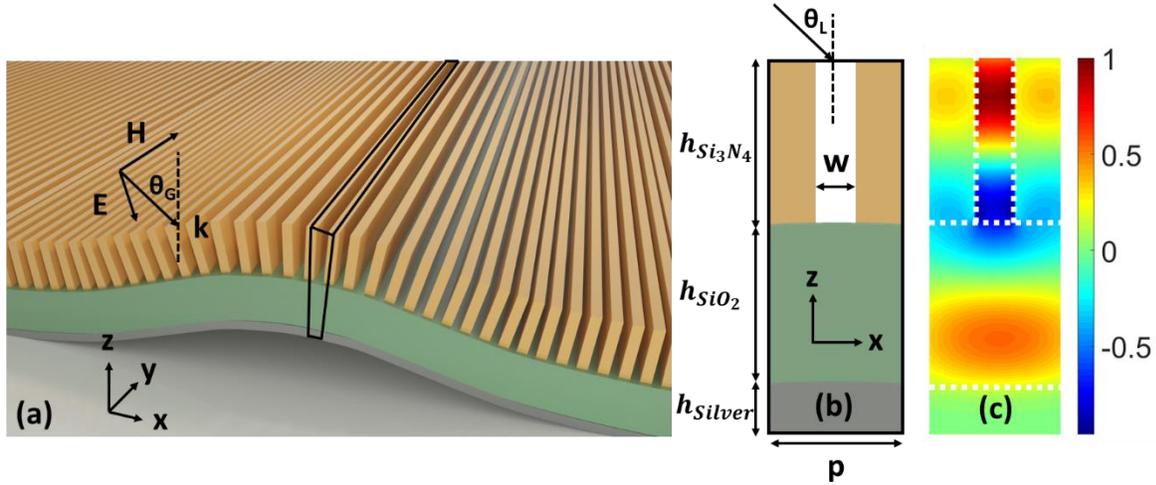

Fig. 2. (a) Schematic of the carpet cloaking system. The incident wave is TM polarized with the magnetic field H along the y direction and with an incident angle $\theta_G$ indicated on the left. (b) The considered geometry of the unit cell of the metasurface. The period (p) of the metasurface array is 280 nm, the width (w) of the air gap is ranging from 80 nm to 200 nm. The metal layer is silver with a thickness of 100 nm ($h_{Silver}$), the spacer is $SiO_2$ with thickness 350 nm ($h_{SiO_2}$), and the slot waveguide consists of two ridges of $Si_3N_4$ with thickness 350 nm ($h_{Si_3N_4}$) separated with an air gap. (c) Normalized electric field at 409 THz with w = 80 nm and $\theta_L$ = 45°.

The design flowchart is as shown in Fig. 3, from the bump geometry z(x) [figure 3(a)], we compute the local incident angle $\theta_L(x)$ [figure 3(b)], and then the phase distribution $\Phi_s(x, f)$ from Eq. 3. Last, for each position x and local incident angle $\theta_L(x)$, we optimize the width to minimize the square of phase error between simulation and the required target phase. Figure 3(c) represents the fitting result at the center of the metasurface (x = 0). The maximum error is less than 20° and it happens at about 384 THz. Figure 3(d) is the fitted width with position. The width is not symmetric with the center of the bump (x=0) because the local incident angle is not symmetric with respect to the center of the bump (x=0).

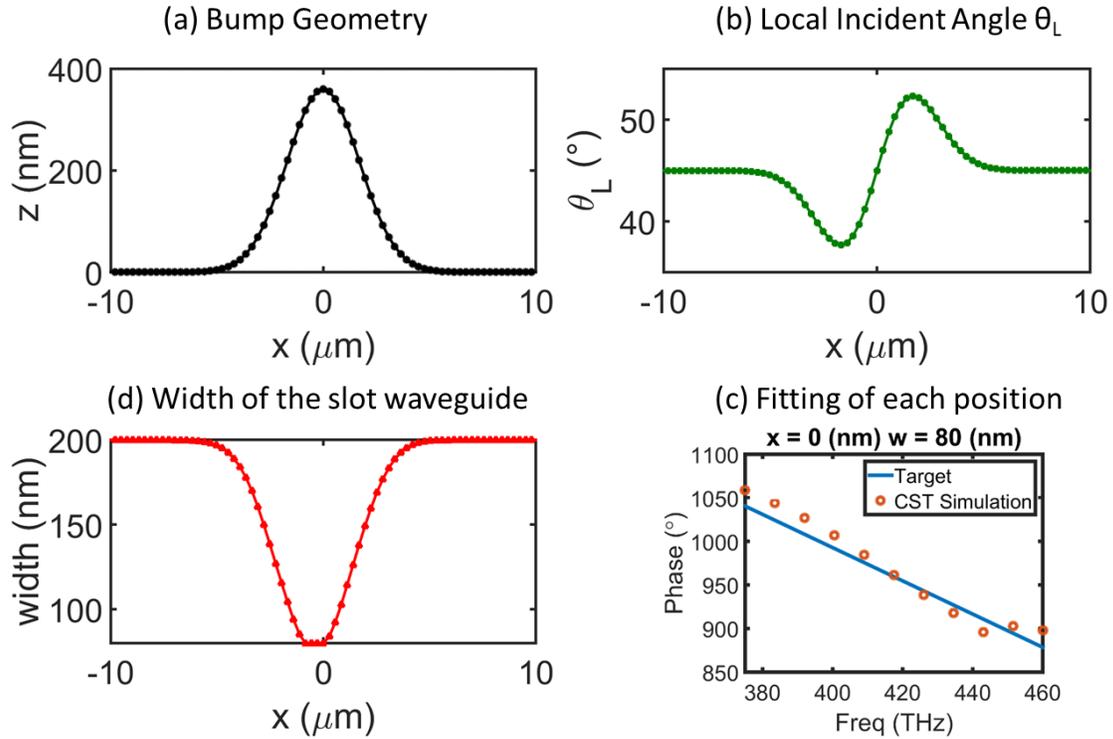

Fig. 3. The design flowchart. (a) Bump geometry as function of the position (b) Local incident angle as function of the position. (c) The fitting result at x=0 (center of the bump). For each position, we optimized the width of the slot waveguide in order to minimize the square of phase error between simulation and the required phase (target). (d) Width of the slot waveguide as function of the position.

**Results**

Figure 4 presents the fitting results from the left of the metasurface to the right. The blue line is the target phase and the orange circles are simulation results. The error increase at the center position. We also noticed that around the frequencies of 390 THz and 430 THz (see (d) & (e)), we have more error (around 25°). It is important to mention that although the target phase is symmetric with respect to the center of the bump(x=0), the width of the slot waveguide is not symmetric with respect to the center of the bump (x=0) at oblique incidence. For example, the local incident angle is 37.67° at x = -1.68 μm but it is 52.33° at x = 1.68 μm.

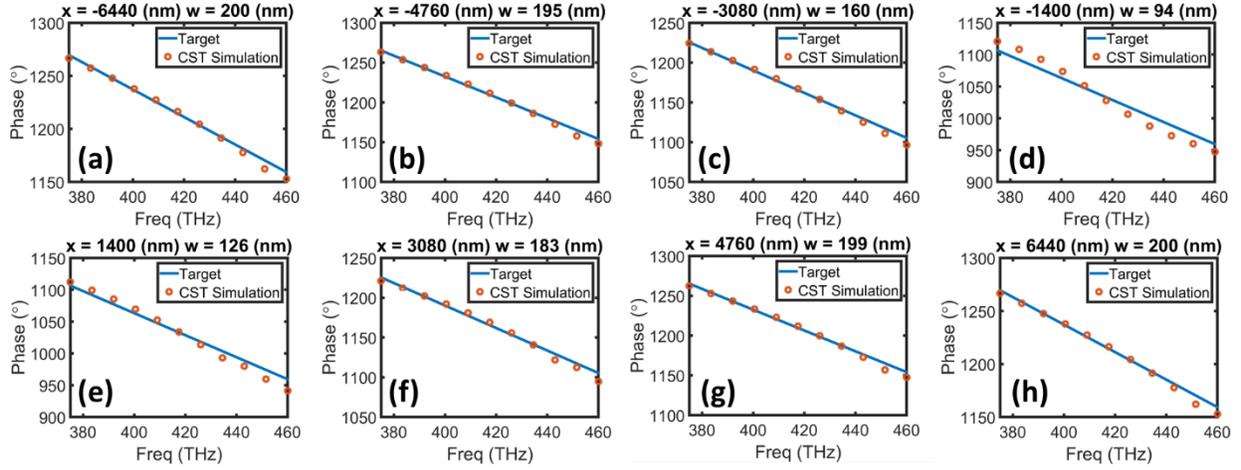

Fig. 4. Fitting results for different positions (a) x = -6440 nm (b) x = -4760 um (c) x = -3080 nm (d) x = -1400 nm (e) x = 1400 nm (f) x = 3080 nm (g) x = 4760 nm (h) x = 6440 nm.

In order to simulate the entire structure, we used the transient solver in CST Microwave Studio. Open boundary conditions are applied to x direction and +z direction and periodic boundary condition is applied to y-direction to mimic invariance along the y-direction. The structure is illuminated at oblique incidence ($\theta_G$=45°) from the left with a TM polarized Gaussian beam. Figure 5 shows the real part of electric field patterns. The columns present different frequencies (375 THz, 392 THz, 409 THz, 426 THz, 443 THz, 460 THz) and the rows present different structures such as the ground plane (a, d, g, j, m, p), the Gaussian-shaped bump (b, e, h, k, n, q), and the Gaussian-shaped bump covered by the designed metasurface (c, f, i, l, o, r). As seen, the wavefronts of reflected waves is flat for the ground plane cases. However, the reflected wavefronts are bent and curved due to Gaussian-shaped bump. After we cover the bump with the metasurface, one can observe that the distortions are corrected, and the reflected wavefronts become flat again. Hence, the designed metasurfaces can cloak the bump well in the broadband range from 375 THz (800 nm) to 460 THz (650 nm).

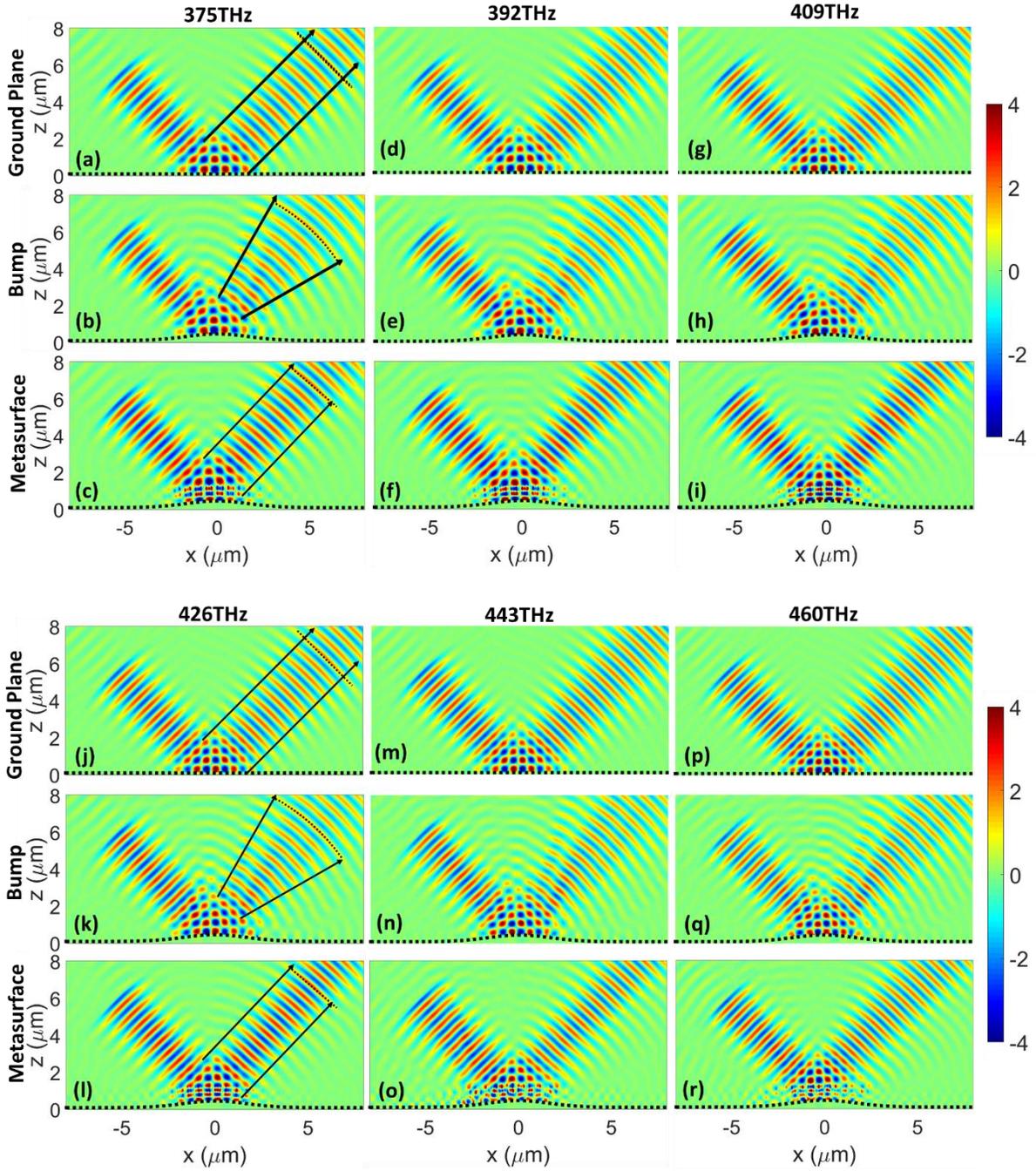

Fig. 5. The real part of electric E field for the ground plane, bump, and metasurface at different frequencies from 375 THz to 460 THz. The columns present different frequencies (375 THz, 392 THz, 409 THz, 426 THz, 443 THz, 460 THz) and the rows present different structures such as the ground plane (a, d, g, j, m, p), the Gaussian-shaped bump (b, e, h, k, n, q), and the Gaussian-shaped bump covered by the designed metasurface (c, f, i, l, o, r).

**Conclusion**

We have proposed and numerically demonstrated a new and simple design for broadband metasurfaces carpet cloaking with linear polarization at visible wavelengths from 650 nm to 800 nm. The design consists of two high index ridges ($Si_3N_4$), separated by a narrow low index gap (air). Due to such refractive index discontinuity forming the slot waveguide, this structure allows one of the propagating modes to confine its energy within the slot region. The advantage of using slot waveguide as a unit cell is to mitigate materials dispersion by confining light in the subwavelength air gap region. Such approach enables broadband operation and paves the way to numerous applications based on broadband metasurfaces/high contrast gratings that were previously unachievable.